\documentclass[twocolumn,amsmath,amssymb]{revtex4}

\def\gsim{ \lower .75ex \hbox{$\sim$} \llap{\raise .27ex \hbox{$>$}} }

\def\lsim{ \lower .75ex \hbox{$\sim$} \llap{\raise .27ex \hbox{$<$}} }

\usepackage{graphicx}
\usepackage{dcolumn}
\usepackage{bm}
\input epsf
\def\gsim{ \lower .75ex \hbox{$\sim$} \llap{\raise .27ex \hbox{$>$}} }
\def\lsim{ \lower .75ex \hbox{$\sim$} \llap{\raise .27ex \hbox{$<$}} }

\usepackage{graphicx}
\usepackage{dcolumn}
\usepackage{bm}

\newcommand{\be}{\begin{equation}}
\newcommand{\ee}{\end{equation}}
\newcommand{\bea}{\begin{eqnarray}}
\newcommand{\eea}{\end{eqnarray}}

\def\e{\epsilon}

\def\t{\tau}

\begin{document}

\title{Dynamical Selection of the Primordial Density Fluctuation
Amplitude}

\author{Jean-Luc Lehners$^{1,2}$ and Paul J. Steinhardt$^{2,3}$}
\affiliation{$^1$ Perimeter Institute for Theoretical Physics, 31 Caroline St N, Waterloo, Ontario N2L 2Y5, Canada \\ $^2$ Princeton Center for Theoretical Science,
Princeton University, Princeton, NJ 08544 USA \\ $^3$ Joseph Henry
Laboratory, Jadwin Hall, Princeton University, Princeton, NJ 08544 USA
}

\begin{abstract}
In inflationary models, the predicted amplitude of primordial density
perturbations $Q$ is much larger than the observed value ($\sim 10^{-
5}$) for natural choices of parameters. To explain the requisite
exponential fine-tuning, {\it anthropic selection} is often invoked,
especially in cases where microphysics is expected to produce a
complex energy landscape.  By contrast, we find examples of ekpyrotic
models based on heterotic M-theory for which {\it dynamical selection}
naturally favors the observed value of $Q$.
\end{abstract}

\maketitle

According to our current understanding, all structure in the universe
originated from primordial density fluctuations. Roughly speaking,
primordial overdense regions acted as seeds that subsequently
underwent gravitational collapse leading to stars and galaxies, while
underdense regions emptied out to form the currently observed voids.
As the COBE \cite{Smoot:1992td} and WMAP \cite{Komatsu:2010fb} satellite experiments have spectacularly
demonstrated, these seeds were already present instants after the big
bang. Inflationary \cite{Baumann:2009ds} and ekpyrotic \cite{Lehners:2008vx} models both provide mechanisms for
generating these density fluctuations by amplifying quantum
fluctuations - in the case of inflation, during a phase of accelerated
expansion after the bang, and, in the case of ekpyrosis, during a
phase of slow contraction before the bang.  Both inflation and
ekpyrosis can be modeled by a scalar field $\phi$ with potential
$V(\phi)$; in the case of inflation the potential is required to be
flat and positive, while for ekpyrosis it must be steep and negative.
In both models, the shape of the potential determines the spectrum of
the fluctuations as well as their
amplitude. It is the latter that we focus on in this work.

In the case of inflation, it is known that the predicted
amplitude of primordial density perturbations $Q$ is much larger than
the observed value ($\sim 10^{-5}$) for natural choices of
parameters \cite{Baumann:2009ds}. For example, for a single field undergoing slow-roll inflation, the number of e-folds of
inflation remaining $N$ and the density fluctuation amplitude (comoving curvature perturbation) $Q$ as a
function of $\phi$ are:
\begin{equation}
N \approx \frac{V}{V_{,\phi \phi}} \;\; {\rm and} \;\; Q \approx
\frac{H^2}{\dot{\phi}}\approx \frac{V^{3/2}}{V_{,\phi}}.
\label{constraints}
\end{equation}
Here $H$ denotes the Hubble rate and we work in reduced Planck units.
To judge the fine-tuning, it is useful to introduce dimensionless
parameters, such as $\alpha \equiv V_{,\phi \phi}^2/(V_{,\phi})^{4/3}$ and
$\beta\equiv V_{,\phi \phi}^2/V$, which are ${\cal O} (1)$ if the
potential is not fine-tuned. For example, for a power-law potential
$V(\phi) = \lambda \phi^4$, $\alpha \approx 23 \lambda^{2/3}$ and
$\beta=144 \lambda$. The two relations in Eq.~(\ref{constraints})
combine to give
\begin{equation}
\alpha  =  Q^{4/3}/N^2.
\end{equation}
Obtaining the observed value of $Q,$
corresponding to $N \sim 60$, requires $\alpha =
{\cal O} (10^{-10}).$
We are not concerned here by the number of orders of magnitude of tuning, which depends on the precise definition of our parameter;
the important point is the trend that larger $Q$
requires less fine-tuning, and so is parametrically favored.
Alternatively, the same relation (or similar ones) shows that, for fixed tuning, the number of e-folds of inflation grows with $Q$; so larger
$Q$ produces universes that are exponentially greater in volume.
Furthermore, larger $Q$ means more structure which means more
galaxies, stars and habitable spaces (up to a certain limiting value
of $Q$ perhaps of order $10^{-2}-10^{-3},$ as described below).  In other words, whether
viewed
parametrically, volumetrically or structure-wise, nothing favors the
observed value of $Q$. For more complicated models, like hybrid inflation, there are more parameters, fields and dimensionless ratios to consider, but it remains the case that nothing favors the observed $Q.$ {\it In short, bad inflation --
producing lots of volume with the wrong cosmic properties -- is more
likely than good inflation}!

A longstanding hope has been to identify microphysics that fixes
parameters uniquely that generate the observed $Q$ and no other. To
date, no accepted microphysics is known that has this property,
although we do not exclude its possibility. (There are cases where tuning is technically natural -- parameters maintain their tuned values after quantum corrections; however, technical naturalness does not, by itself, explain why $Q$ has the observed value.) An alternative is that
microphysics (such as string theory) leads to a complex energy
landscape with a discretuum of possible values of inflationary
parameters.  The parameters that vary in this way are said to
``scan.''
In this case, the observed value of $Q$ does not seem to be preferred in any way, and generically larger $Q$ is favored.
Anthropic selection is then
invoked to resolve the bad inflation problem.  An anthropic
upper limit of $Q \lesssim 3 \cdot 10^{-4}$ has been suggested based
on requiring planetary orbits, with a radius similar to Earth's radius
around the associated star, not to be disrupted within a timeframe of
a billion years \cite{Tegmark:1997in}. However, it is not clear how seriously to take this
bound.  It is based on requiring typical planets to have strong
similarities to the Earth, which does not seem essential for life, and
it does not weigh the fact that larger $Q$ is much more likely (for
the reasons described above).  In fact, until $Q$ was actually
measured, values of $Q$ that are an order of magnitude or two larger
appeared compatible with our own observable universe. If we take this
as a more conservative anthropic bound, then inflation favors these
larger values and the anthropic selection does not explain the
observed value.

In this paper, we consider the same issue for ekpyrotic/cyclic models
in which density fluctuations are generated prior to the big bang and
inflation is avoided. Although versions of these models can be
constructed from conventional 4d scalar fields and potentials, here we
will return to the setting that originally motivated the idea:
heterotic M-theory \cite{Khoury:2001wf}. In this case, the big
crunch/big bang transition corresponds to the collision and bounce
between branes (orbifold planes) along an extra spatial direction, and
smoothing, flattening, and density perturbation generation occur as
the two branes approach one another before the bounce. The colliding brane picture is currently unproven. However, for the purposes of this study, we will assume it is viable and demonstrate a significant cosmological consequence; and to the extent to which the result is intriguing, it provides motivation for exploring this picture further to determine its validity.

We will show
that, in the colliding brane picture, the correct value of $Q$ results from non-anthropic,
dynamical selection even if there is a complex landscape where
parameters scan.
More precisely, there is an upper bound on $Q$ which depends explicitly on the value of Newton's constant and implicitly on the gauge coupling constants. Then, given the observed values for these physical constants, the upper bound on $Q,$ which is dynamically selected, agrees precisely with its observed value.

In heterotic M-theory \cite{Lukas:1998yy}, six small
extra dimensions are wrapped in a Calabi-Yau manifold, and our brane
is separated from a parallel one by an additional, eleventh, extra
dimension. In the 4d effective field theoretic description, the
ekpyrotic phase is modeled by a steep, negative potential of the form
$V=-V_0 e^{-\sqrt{2\e}\phi},$ where $\e \approx 10^2$ is called the
fast-roll parameter. During this phase, our universe slowly contracts
in the 4d effective description according to the scaling solution \be
a=(-\tau)^{1/\e}, \quad \frac{\mathrm{d}\phi}{\mathrm{d}\tau} = \sqrt{2}/({\sqrt{\e}\,\t}),
\quad V = -\e {\cal H}^2, \ee
using a conformal time coordinate $\t$ which is negative before the
big bang, and approaches zero at the collision between branes (the big crunch/big bang
transition). The conformal Hubble rate is written as ${\cal H} \equiv (\frac{\mathrm{d}a}{\mathrm{d}\tau})/a.$
During the contraction phase, the
scalar $\phi$ rolls down the potential to the maximal depth $-V_{ek},$
at which point the potential increases and approaches zero. At that
minimum, which we denote by $\t_k,$ the ``kinetic'' phase begins.
During this phase, the energy density is dominated by the kinetic
energy of $\phi,$ and the corresponding solution is \be a = (-
\t)^{1/2}, \qquad \frac{\mathrm{d}\phi}{\mathrm{d}\tau}= {\sqrt{3}}/({\sqrt{2}\,\t}). \ee
Matching
the conformal Hubble rates in the two phases, we find \be -2\t_k =
\sqrt{{\e}/{V_{ek}}}.\label{matchingHubble}\ee From the 11d
braneworld point of view, the two parallel branes, which are at a
distance $d_{11}$ at the time $\t_k,$ approach each other and collide
at $\t=0$. Conventionally, the brane velocities relative to the center
of mass are denoted by $y_0$, and so, to leading order in $y_0,$ we
have that \be d_{11} = 2 y_0 |\t_k|.\ee (The inter-brane velocity is
actually  $\tanh(2y_0),$ but our approximation is valid since the
branes move at non-relativistic speeds in the regime of interest.)
Combined with (\ref{matchingHubble}), we can obtain an expression for
the potential minimum in terms of the collision velocity \be V_{ek} =
{\e y_0^2}/{d_{11}^2}. \label{PotentialVelocity}\ee Since
the fluctuation amplitude $Q$ goes as the square root of
the potential minimum, $Q \propto V_{ek}^{1/2},$ we have \be Q \propto
y_0. \ee This simple result is at the core of our argument. The brane
collision velocity cannot be arbitrarily high - semi-classical studies
of the brane collision have shown that radiation and matter are
produced at the brane collision, in quantities increasing with $y_0$
\cite{Turok:2004gb}. Once the collision velocity reaches relativistic
values, {\it i.e.} for $y_0 \gtrsim 0.1,$ the matter density produced
at the collision reaches the Hagedorn density, and the branes re-
collapse rapidly under gravity. That is, the branes ``stick'' together
for relativistic collision speeds, and no bounce or expanding universe
follows. Thus, there is an upper limit on $y_0,$ and correspondingly,
{\it an upper limit on the primordial density fluctuation amplitude}
$Q!$

It is important to realize that this upper limit is not inferred because we have reached the limit of validity of the 4d effective field theory, but because a catastrophic event, namely the collapse of the brane universe, occurs when the limit is surpassed.
The exact numerical value of this upper limit $Q_{max}$ will depend on
the details of the ekpyrotic or cyclic model under study. We will
illustrate our argument with one of the best-understood models to
date, namely the heterotic M-theory colliding branes solution
\cite{Lehners:2006pu}, which incorporates the entropic mechanism
\cite{Lehners:2007ac} for producing density perturbations (for a
review, see \cite{Lehners:2008vx}). In this model, the amplitude of
the (nearly scale-invariant) density fluctuations is given by \be Q^2
\approx \frac{\e V_{ek}}{10^3}  \approx
\frac{y_0^2 \epsilon^2}{10^3 d_{11}^2}, \label{fluctamplitude2}\ee
where we have used (\ref{PotentialVelocity}). As stated earlier, the
upper limit on $y_0$ is about $0.1,$ while $\e\approx 10^2.$ So it
remains to determine the appropriate value of $d_{11}.$ In the
colliding branes solution, the 11-dimensional distance between the
branes was calculated in Eq. (3.38) of Ref. \cite{Lehners:2006pu} to
be \bea
d &=& \begin{cases} |2y_0\t| & \mbox{for } |\t| \leq |\t_{bntb}| \\
d_{11} & \mbox{for }
|\t| \geq |\t_{bntb}|. \label{distance}
\end{cases}\eea A few clarifications are in order: the above
expression verifies that close to the collision, the inter-brane
velocity is indeed $2 y_0$ (in \cite{Lehners:2006pu}, $d$ was only
evaluated in this regime). Before the time $\t_{bntb},$ the inter-
brane distance is fixed (to leading order in $y_0$) at the constant
value $d_{11}.$ The time $\t_{bntb}$ marks a
special event, namely the moment when, from the higher-dimensional
point of view, the negative-tension brane bounces off a naked
singularity in the 11-dimensional bulk spacetime (the details are
unimportant for us here, but can be found in
\cite{Lehners:2006pu,Lehners:2007nb,Lehners:2006ir}).
What is important is
that, throughout the ekpyrotic phase, the inter-brane distance is
fixed at the value $d_{11},$ and this implies that, in a cyclic
context, we can identify $d_{11}$ with its current value.
As first discussed by Ho\v{r}ava and Witten \cite{Horava:1995qa,Witten:1996mz},
the volume of the internal space determines the gauge coupling constant of grand unification, while both the volume and $d_{11}$ determine Newton's
constant $G_N$. This allows us to fix the value of $d_{11}$ at the phenomenologically
required value of (in reduced Planck units)
\cite{Banks:1996ss}
\be d_{11} \approx 10^{3.5}.\ee  We
are finally in a position to evaluate our bound on the fluctuation
amplitude $Q:$ since by definition $|\t_{bntb}|<|\t_k|,$ using
(\ref{fluctamplitude2}) we obtain the upper limit \be Q_{max} \approx
10^{-4},\ee which shows that this model yields an upper bound on $Q$
that is very close to the observed value.
Moreover, we note that,
since there must be some time between the moment when the potential
minimum is reached and the bounce of the negative-tension brane, a more
realistic approximation would be $|\t_k| \approx (2-3) |\t_{bntb}|;$
with this improved estimate, the maximal value is $Q_{max} \approx
10^{-4.5},$ which, considering the level of approximation we are
working at, coincides with the observed value.

Hence, considering only the formation of large-scale structure and as argued above for the case of inflationary cosmology, the
bounce that produces the highest possible $Q$ is preferred because it
produces more galaxies, stars, etc.  {\it
Imposing the prerequisite  that
there be a bounce, this dynamical selection principle naturally
picks out the observed amplitude of primordial density fluctuations.}
One could argue that we were successful numerically because we imposed the constraint that Newton's constant $G_N$ and the gauge couplings have the observed values. If these couplings also were to scan, then the preferred $Q$ would be different from one landscape minimum to the next. This does not diminish the significance of our result, though. In this case, we amend our conclusion to read that the ekpyrotic model predicts a {\it correlation} on the landscape between $G_N$, the gauge couplings and $Q:$ that is, for minima that have $G_N$ and gauge couplings like those we observe, $Q$ is predicted to be ${\cal O}(10^{-4}-10^{-5}).$

In order to complete our argument, we will now show what other effects
the change in $V_{ek}$ causes to the ekpyrotic or cyclic universe, in
particular in terms of volume and tuning, and that these changes do
not alter our result.
The radiation energy density at the bang is expected to be
proportional to the energy of the collision, \be H_r^2 \approx T_r^4
\propto y_0^2, \label{collisionenergy}\ee where $H_r$ and $T_r$ denote
the Hubble rate and temperature at the onset of radiation domination.
However, as can be inferred from Eq. (\ref{PotentialVelocity}), the
energy of the collision is proportional to $V_{ek}.$ Moreover, the
expansion of the universe since the big bang is given by $(T_r/T_0),$
where $T_0$ denotes the current temperature of the cosmic microwave
background. Thus, in the context of the ekpyrotic model, where we
assume a one-time ekpyrotic phase followed by a brane collision/big
bang, we can conclude that, by varying $V_{ek},$ the volume of the
universe changes as $V_{ek}^{3/4}.$   Therefore increasing $V_{ek},$
and thus increasing $Q,$ implies that a larger universe is produced;
that is, higher $Q$ (but consistent with a bounce) is preferred
volumetrically.

For the cyclic universe, the analysis is more involved. The
main feature of the cyclic universe is that it provides a resolution
of the problem of initial conditions in that it dynamically selects
the right initial conditions cycle after cycle. However, in the two-
field cyclic universe which we are discussing here, during the
ekpyrotic phase there is an instability associated with the field
direction transverse to $\phi$ \cite{Lehners:2007ac,Tolley:2007nq}.
This instability has the consequence that only a small fraction of the
universe makes it from one cycle to the next, but this is more than
compensated for by the net expansion over the course of one cycle, if
there is a sufficient amount of dark energy expansion
\cite{Lehners:2008qe,Lehners:2009eg}. Then, overall, the volume of
habitable space grows from cycle to cycle. Let us now look at the
details:

The ekpyrotic phase lasts for a number $ N_{ek} \equiv \frac{1}{2}
\ln ({V_{ek}}/{V_0}) $ of e-folds of slow contraction (with $V_0
\approx T_0^4$ denoting the current dark energy scale). During this
phase, the universe contracts by a negligible total amount, but a
fraction $e^{-3N_{ek}}$ of the volume of the universe is thrown off
the ideal cyclic trajectory due to the above-mentioned instability.
These regions are likely to experience a fatal chaotic
mixmaster big crunch and never bounce; they are irrelevant for galaxy
formation.

The net increase in volume per cycle is given by
\cite{Erickson:2006wc} \be \Bigg(\frac{V_{ek}^{1/4}}{T_r}\Bigg)^{2}
\Bigg(\frac{T_r}{T_0}\Bigg)^3 e^{3N_{de}},\ee where the successive factors
denote the expansion during the kinetic phase, radiation and
matter domination and during the dark energy phase (lasting $N_{de}$ e-folds). The scalar field $\phi$ reaches $-\infty$ at
the big crunch/big bang transition \cite{Steinhardt:2001st}, and rebounds
in a matter of instants to the value near what it is today.  To be
precise, the field is stopped by the Hubble damping to radiation at
$\phi_{stop} \sim \ln (y_0/H_r),$ before turning back.
Note that, the bigger the collision velocity is, the further away from $-
\infty$ the field rebounds; but the greater is the radiation damping,
the quicker the field comes to a halt. From the relation
(\ref{collisionenergy}) above, the net result is that $\phi_{stop}$ is
unchanged. This also means that the number of e-folds of dark energy
$N_{de} \approx e^{\sqrt{\e}\phi_{stop}}/\e$ is unchanged
and independent of $V_{ek}.$

Putting these relations together, we conclude that the fractional
increase $F$ in habitable volume per cycle is given by
$F=(T_r/T_0) e^{-2N_{ek}+3N_{de}}.$
Since we are assuming that $T_r \propto V_{ek}^{1/4},$ this can be
rewritten as \be F=(e^{-\frac{1}{2}N_{ek}+N_{de}})^3.\ee Incidentally
this shows that the requirement for the two-field cyclic universe to
be sustainable is $N_{de} \geq \frac{1}{2}N_{ek}.$ Since $N_{de}$ is
fixed, we have that \be F \propto V_{ek}^{-3/4}. \ee Thus, increasing
the depth of the potential actually produces a smaller universe.
However, since an increase in $V_{ek}$ also corresponds to an increase
in $Q \propto V_{ek}^{1/2},$ this smaller universe will contain
exponentially
more structure.
If we use Press-Schechter \cite{Press:1973iz} as an approximation,
then the probability for a region to have a density contrast $\delta$
is proportional to $(1/Q)\,e^{-\delta^2/Q^2}.$ In our
(region of the) universe, we have $Q^2 \approx 10^{-9}.$ Hence, in the
context of the cyclic universe, decreasing the value of $V_{ek}^{1/4}$
by a factor of $10$ produces a universe that is $10^3$ times bigger
but produces a value of $Q$ that is 100 times smaller;  the net effect
is to decrease the probably of galaxies (and life) exponentially.  So,
as in the simple ekpyrotic case, the largest $V_{ek}$ compatible with
a bounce is preferred.

In sum, we have pointed out that the conventional inflationary picture
suffers from the bad inflation problem: regions with $Q$ larger than
observed are preferred.  They require less fine-tuning, produce more
structure and, in most cases, generate more volume.  In cases with a
complex energy landscape in which inflationary parameters scan, the
problem is particularly acute -- our observed universe is highly unlikely.  Anthropic selection reduces but does
not resolve the problem.  As an alternative, we have presented an
example of an ekpyrotic/cyclic model where the observed value of $Q$
is selected dynamically, even in cases where there is a complex energy
landscape. Being able to do this when parameters scan is novel.  This
raises the interesting question, which we leave to future work, of
what other predictions can be extracted from this framework.  Note
that we do not claim that the model presented here offers the unique
explanation of the value of $Q;$ nor do we argue that dynamical
selection with a complex energy landscape is impossible for
inflationary models.  Rather we want to provide an existence proof
that a dynamical selection is possible and suggest that such an
explanation should be added as an advantageous criterion in judging models of the
early universe.

\noindent {\it Acknowledgements:} We thank D. Spergel for useful remarks  and A. Albrecht, D. Baumann and J. Khoury  for their review and comments on the original draft. This work is supported in part by Department of Energy Grant DE-FG02-91ER40671.

\end{document}